\documentclass[%
aip,
jcp,%
amsmath,amssymb,
reprint
]{revtex4-1}
\usepackage[latin1]{inputenc}

\usepackage{amsmath}
\usepackage{amsfonts}
\usepackage{amssymb,bm}
\usepackage{graphicx}
\usepackage{physics}
\usepackage{upgreek}
\usepackage[outercaption]{sidecap}   
\usepackage{color}
\usepackage{newtxtext}
\usepackage{newtxmath}
\usepackage[version=4]{mhchem}

\DeclareMathAlphabet{\pazocal}{OMS}{zplm}{m}{n}

\newcommand{\pL}{\ensuremath{\pazocal{L}}}

\newcommand{\pR}{\ensuremath{\pazocal{R}}}

\newcommand{\el}{\ensuremath{\mathrm{e}}}

\newcommand{\sing}{\ensuremath{\mathrm{S}}}
\newcommand{\trip}{\ensuremath{\mathrm{T}}}

\newcommand{\sX}{\ensuremath{\mathsf{X}}}
\renewcommand{\op}[1]{\ensuremath{\hat{#1}}}

\usepackage{booktabs}

\begin{document}
	
\title{Spin relaxation in radical pairs from the stochastic Schr\"odinger equation}
\author{Thomas P. Fay}
\email{tom.patrick.fay@gmail.com}
\affiliation{Department of Chemistry, University of Oxford, Physical and Theoretical Chemistry Laboratory, South Parks Road, Oxford, OX1 3QZ, UK}
\author{Lachlan P. Lindoy}
\affiliation{Department of Chemistry, University of Oxford, Physical and Theoretical Chemistry Laboratory, South Parks Road, Oxford, OX1 3QZ, UK}
\author{David E. Manolopoulos}
\affiliation{Department of Chemistry, University of Oxford, Physical and Theoretical Chemistry Laboratory, South Parks Road, Oxford, OX1 3QZ, UK}

\begin{abstract}
We show that the stochastic Schr\"odinger equation (SSE) provides an ideal way to simulate the quantum mechanical spin dynamics of radical pairs. Electron spin relaxation effects arising from fluctuations in the spin Hamiltonian are straightforward to include in this approach, and their treatment can be combined with a highly efficient stochastic evaluation of the trace over nuclear spin states that is required to compute experimental observables. These features are illustrated in example applications to a flavin-tryptophan radical pair of interest in avian magnetoreception, and to a problem involving spin-selective radical pair recombination along a molecular wire. In the first of these examples, the SSE is shown to be both more efficient and more widely applicable than a recent stochastic implementation of the Lindblad equation, which only provides a valid treatment of relaxation in the extreme-narrowing limit. In the second, the exact SSE results are used to assess the accuracy of a recently-proposed combination of Nakajima-Zwanzig theory for the spin relaxation and Schulten-Wolynes theory for the spin dynamics, which is applicable to radical pairs with many more nuclear spins. An appendix analyses the efficiency of trace sampling in some detail, highlighting the particular advantages of sampling with $SU(N)$ coherent states.	
\end{abstract}

\maketitle
	
\section{Introduction}

Spin relaxation effects can influence the qualitative and quantitative behaviour of many spin chemical systems, the archetype of which is the radical pair.\cite{Rodgers2009,Steiner1989} A radical pair consists of two radicals, undergoing spin-selective reactions, with unpaired electron spins which are coupled to each other and to a set of nuclear spins. These are encountered in many contexts, including light-harvesting molecular devices,\cite{Wasielewski2006, Scott2009,Scott2011} organic LEDs,\cite{Geng2016,Geng2016a,Hagi2019}, molecular qubits,\cite{Sun2014,Rugg2017,Wu2018}  and various biological processes,\cite{Grissom1995,Brocklehurst2002,Prakash2005,Hoff1981,Biskup2009,El-Esawi2017} potentially including the magnetic compass sense of migratory birds.\cite{Rodgers2009a,Hore2016,Wiltschko2019} Accurate modelling of the radical pair spin dynamics requires quantum mechanical calculations which include the effects of spin relaxation.\cite{Steiner1989,Atherton1993,Goldman2001} Efficiently performing such calculations on realistic models of radical pairs presents a significant challenge which we shall address in this paper.

Spin relaxation processes present a particular challenge in the quantum mechanical modelling. Relaxation is caused by stochastic fluctuations in spin interactions in radical systems, resulting from thermal motion of the molecules. For a given model for these fluctuations, a Stochastic Liouville equation (SLE) can be derived,\cite{Kubo1969,Freed1971,Vega1975,Lau2010, Vega1975, Pedersen1973, Pedersen1994}  which gives the exact quantum spin dynamics of the radical pair spin system.\cite{footnote1} The disadvantage of this approach is the large computational cost of working with an extended spin density operator that includes the stochastically fluctuating variables, which limits the application of the SLE to unrealistically small spin systems. To circumvent this problem, various approximate theories can be applied, such as Bloch-Redfield-Wangsness\cite{Wangsness1953,Redfield1965,Goldman2001} theory and Nakajima-Zwanzig theory,\cite{Nakajima1958,Zwanzig1960} and the Lindblad equation that can be derived as a further approximation to these theories. However these approaches still involve working with the full spin density operator, so calculations are still restricted to relatively small spin systems. 

Recently Keens \& Kattnig have demonstrated that a Monte Carlo wavefunction approach can be modified to find solutions to the Lindblad equation for recombining radicals,\cite{Keens2020} and that this approach can be combined with a coherent state sampling scheme introduced by us to reduce the computational cost of spin dynamics calculations in systems without spin relaxation.\cite{Lewis2016} This makes spin dynamics simulations including relaxation feasible for realistic models of radical pairs, {\em provided} the relaxation can be treated accurately with a Lindblad equation. 

We ourselves have recently shown\cite{Fay2019a} how to rigorously combine a perturbative Nakajima-Zwanzig treatment of the spin relaxation with the Schulten-Wolynes semiclassical approximation\cite{Schulten1978} to the nuclear spins. This provides a computationally efficient method for modelling spin dynamics in radical pair systems, and it is expected to be reasonably accurate for radical pairs with short lifetimes or strong electron spin interactions. However, this method is only expected to work if the fluctuations in the spin Hamiltonian causing relaxation can be treated perturbatively, and if there are a relatively large number of hyperfine coupled nuclear spins. It should also be mentioned that kinetic master equations,\cite{Hayashi1984,Steiner2018,Fay2019,Mims2019,Riese2020} and improved semiclassical methods\cite{Manolopoulos2013,Lewis2014} can be employed to include relaxation effects in spin dynamics calculations, but these methods also have their limitations. In fact, all existing techniques for modelling relaxation in radical pairs have their shortcomings, which motivates the development of a more accurate and generally applicable method for treating radical pairs with a realistic number of nuclear spins. 

In this paper, we propose applying the stochastic Schr\"odinger equation (SSE) to this problem. This provides an exact method for modelling relaxation which is significantly more efficient for large spin systems than density operator based methods such as the Lindblad equation and the SLE. In the SSE approach, the stochastic fluctuations in the spin interactions are directly included in the spin state dynamics. This can be viewed as a way to extend the Keens \& Kattnig Monte Carlo wavefunction method\cite{Keens2020} to treat relaxation beyond the Lindblad approximation, an idea which we have already suggested in several papers.\cite{Lewis2016,Fay2017,Fay2019a} Indeed, the direct use of stochastic fluctuations to model relaxation in spin dynamics is not new: it has previously been applied, for example, in the context of simulating both EPR spectra\cite{Saunders1968,Robinson1992,Sezer2008} and radical pair recombination.\cite{Nielsen2019} However, these previous applications have been limited to relatively small spin systems. Here we shall show how efficient trace sampling techniques can seamlessly be combined with the SSE to facilitate the treatment of larger systems.

We begin in Section \ref{theory-sec} by outlining the theory of radical pair recombination reactions, and describing how to evaluate quantum mechanical expressions for observables using the SSE in combination with trace sampling. In Section \ref{results-sec} we then apply this method to a set of test problems. Firstly we consider the effect of random field relaxation on some model FAD$^{\bullet -}$ -- W$^{\bullet +}$ and FAD$^{\bullet -}$ -- Z$^{\bullet}$ radical pairs of relevance to the avian magnetoreception problem, extending the model systems treated by Keens \& Kattnig in Ref.~\onlinecite{Keens2020} to include non-Markovian effects which cannot be captured with the Lindblad equation. Secondly we simulate the spin dynamics of \ce{DMJ^{$\bullet+$}-An-Ph_$\text{n}$-NDI^{$\bullet-$}} \lq\lq molecular wire" radical pairs with a more realistic rotational Brownian motion model of relaxation. We use these simulations to investigate the accuracy of a combination of Nakajima-Zwanig theory and Schulten-Wolynes theory which we have recently used\cite{Fay2019a} to interpret magnetic field effect experiments\cite{Scott2009} on these radical pairs. Our conclusions are drawn in Section~\ref{conc-sec}. 

\section{Theory}\label{theory-sec}

\subsection{Spin dynamics in radical pairs}

The spin density operator $\op{\rho}(t)$ contains all information on the observables of a radical pair spin system. For a radical pair undergoing spin-selective recombination reactions, this evolves according to the Haberkorn master equation,\cite{Haberkorn1976,Ivanov2010,Fay2018}
\begin{align}\label{haber-eq}
	\dv{t}\op{\rho}(t) = -\frac{i}{\hbar}\left[\op{H},\op{\rho}(t)\right] - \left\{\op{K},\op{\rho}(t)\right\}.
\end{align}
Here $\op{H}$ is the effective spin Hamiltonian and $[\op{A},\op{B}]$ is the commutator of $\op{A}$ and $\op{B}$. $\op{K} = (k_\sing/2)\op{P}_\sing + (k_\trip/2)\op{P}_\trip$ is the operator that describes the spin-selective recombination reactions, in which singlet radical pairs recombine at a rate $k_\sing$ and triplet radical pairs at a rate $k_\trip$. $\op{P}_\sing$ and $\op{P}_\trip$ are the projection operators onto the singlet and triplet electronic states of the radical pair, and $\{\op{A},\op{B}\}$ is the anti-commutator of $\op{A}$ and $\op{B}$. 

The initial value of the spin density operator is determined by how the radical pair is formed. If it is formed in a singlet electronic state, the initial density operator will be $\op{\rho}(0) = \op{P}_\sing/Z$, where $Z$ is the dimensionality of the nuclear spin Hilbert space. For simplicity this is the only case we shall consider here, although there is no difficulty in generalising what follows to treat other electronic spin state initial conditions.

The spin Hamiltonian $\op{H}=\op{H}_{12}+\op{H}_1+\op{H}_2$ contains a term $\op{H}_{12}$ which couples the electron spins, and single radical terms $\op{H}_i$ which describe the spin interactions in radicals $i=1\text{ and }2$. The electron spin coupling term is given in general by\cite{Steiner1989,Rodgers2009}
\begin{align}
	\op{H}_{12} = -2J\, \op{\vb{S}}_1\cdot \op{\vb{S}}_2 + \op{\vb{S}}_1\cdot \vb{D} \cdot \op{\vb{S}}_2,
\end{align}
where $\op{\vb{S}}_i$ is the unitless spin operator for the electron spin in radical $i$, $J$ is a scalar coupling constant, and $\vb{D}$ is a dipolar coupling tensor. The single radical terms can be written as\cite{Steiner1989,Rodgers2009}
\begin{align}
	\op{H}_i = \mu_\mathrm{B} \vb{B}\cdot \vb{g}_i \cdot \op{\vb{S}}_i + \sum_{k=1}^{N_k} \op{\vb{I}}_{ik} \cdot \vb{A}_{ik} \cdot \op{\vb{S}}_i.
\end{align}
Here the first term describes the electronic Zeeman interaction, where $\mu_\mathrm{B}$ is the Bohr magneton, $\vb{B}$ is the applied magnetic field strength, and $\vb{g}_i$ is the electron spin $g$-tensor of radical $i$. The second term describes the hyperfine interactions between the electronic and nuclear spins in the radical. $\op{\vb{I}}_{ik}$ is the nuclear spin operator of nucleus $k$ in radical $i$ with spin quantum number $I_{ik}$, and $\vb{A}_{ik}$ is the hyperfine coupling tensor for this nuclear spin, which includes both isotropic Fermi contact and anisotropic dipolar contributions.\cite{Steiner1989,Rodgers2009}

\subsection{Nuclear motion and spin relaxation}

The spin-coupling parameters in the radical pair Hamiltonian all depend on the instantaneous nuclear configuration. This fluctuates due to the thermal motion of the atoms in the radical pair, causing modulation of the spin Hamiltonian (for example, rotational motion modulates anisotropic spin couplings), which leads to spin relaxation. 

This microscopic picture of the origin of spin relaxation can be modelled by introducing a set of stochastically fluctuating variables $\sX(t)$ which describe the molecular motions, and letting the spin Hamiltonian fluctuate in time as a function of these variables: $\op{H}(t) \equiv \op{H}(\sX(t))$.\cite{footnote2} In general, the set of variables $\sX(t)$ can be continuous, for example the Euler angles describing the orientation of a molecule, $\Upomega(t) = (\upalpha(t),\upbeta(t),\upgamma(t))$,\cite{Nicholas2010} or discrete, for example the variables associated with the conformational changes of a radical undergoing stochastic hops between discrete torsional minima.

When the effect of molecular motions on the spin dynamics is modelled in this way, observables must be obtained from the density operator averaged over realisations of $\sX(t)$, which we denote $\ev{\op{\rho}(t)}$. From this, ensemble averaged observables $\ev{O(t)}$ are extracted using the corresponding operator $\op{O}$ as
\begin{align}
	\ev{O(t)}= \Tr[\op{O}\ev{\op{\rho}(t)}],
\end{align}
where $\Tr[\cdots]$ denotes the trace over the spin Hilbert space. For example, the total radical pair survival probability $\ev{1(t)}$ can be calculated using $\op{O} = \op{1}$, the singlet radical pair survival probability $\ev{P_\sing (t)}$ using $\op{O} = \op{P}_\sing$, and the triplet radical pair survival probability $\ev{P_\trip (t)}$ using $\op{O}=\op{P}_\trip$.

Conventionally, $\ev{O(t)}$ is obtained by solving the SLE for the density operator $\op{\rho}(t,\!\sX) = \ev{\op{\rho}(t)\delta(\sX - \sX(t))}$, 
\begin{align}\label{sle-eq}
	\dv{t}\op{\rho}(t,\!\sX) \!=\! -\frac{i}{\hbar}\left[\op{H}(\sX),\op{\rho}(t,\!\sX)\right] \!-\! \left\{\op{K},\op{\rho}(t,\!\sX)\right\} \!+\! \Gamma \op{\rho}(t,\!\sX),
\end{align}
in which $\Gamma$ is an operator on functions of $\sX$ which describes the evolution of the probability density for $\sX$. For example, for free rotational Brownian motion, $\Gamma = -\sum_{\alpha = X,Y,Z} D_\alpha\mathsf{L}_\alpha^2$, where $\mathsf{L}_\alpha$ is the $\alpha$ component of a unitless body-fixed angular momentum operator and $D_\alpha$ is the corresponding component of the body-fixed rotational diffusion tensor.\cite{Nicholas2010} 

In practice, this SLE calculation proceeds by expanding $\op{\rho}(t,\!\sX)$ in some finite set of basis functions of $\sX$, solving Eq.~(5) in this basis, integrating the resulting $\op{\rho}(t,\!\sX)$ over $\sX$ to obtain $\ev{\op{\rho}(t)}$, and then substituting this into Eq.~(4) to obtain $\ev{O(t)}$. However, working in Liouville space becomes very expensive for large spin systems (both in terms of computer time and computer memory), and using a basis set for the configurational ($\sX$) variables simply adds to this expense. It is significantly more efficient to work in Hilbert space, and to treat $\sX(t)$ as a set of stochastic variables, as we shall describe next.

\subsection{Trace sampling and the stochastic Schr\"odinger equation}

When the spin Hamiltonian is time-dependent (i.e., for a given realisation of the fluctuating variables $\sX(t)$), the solution to Eq.~\eqref{haber-eq} can be written as
\begin{align}
	\op{\rho}(t) = \op{U}(t) \op{\rho}(0) \op{U}(t)^\dag,
\end{align}
where the propagator $\op{U}(t)$ is
\begin{align}
	\op{U}(t) = \mathsf{T} \exp[\int_0^t \dd{\tau}\left(-\frac{i}{\hbar}\op{H}(\sX(\tau)) - \op{K}\right)],
\end{align}
in which $\mathsf{T}$ denotes the forwards time-ordering operator. Assuming the singlet initial condition $\op{\rho}(0)=\op{P}_\sing/Z = \ket{\sing}\bra{\sing}/Z$ for simplicity, we can use this propagator to write the ensemble-averaged expectation value of an observable in terms of a nuclear spin state trace, which we denote $\tr[\cdots]$, as
\begin{align}\label{ot-tr-eq}
	\ev{O(t)} = \frac{1}{Z}\ev{\tr[\ev{\op{U}(t)^\dag \op{O}\op{U}(t)}{\sing}]},
\end{align}
where the outer angular brackets denote an average over realisations of the fluctuating variables.\cite{footnote3} All that remains to turn this into a practical expression is to find an efficient way of evaluating the  trace that avoids applying the evolution operator $\hat{U}(t)$ separately to the direct product of $\ket{\sing}$ with each of the $Z$ states that span the nuclear spin Hilbert space.
 
For this, we can exploit some well-established results concerning the stochastic evaluation of quantum mechanical traces.\cite{Weisse2006} In particular, suppose we have a set of normalised nuclear spin states $\ket{\psi(\boldsymbol{\xi})}$, parametrised by a set of real variables $\boldsymbol{\xi}$, with which we can resolve the nuclear spin identity operator as
\begin{align}\label{gen-res-id-eq}
	\op{1} = Z\int \dd{\boldsymbol{\xi}} p(\boldsymbol{\xi}) \dyad{\psi(\boldsymbol{\xi})},
\end{align}
where $p(\boldsymbol{\xi})$ is a normalised probability density for $\boldsymbol{\xi}$. There are many such resolutions of the identity, and we will give some specific examples below. But for now, let us stick with Eq.~(9) for generality. 

With Eq.~\eqref{gen-res-id-eq}, we can re-write the nuclear spin trace as 
\begin{align}
	\tr[\op{A}] = Z\int \dd{\boldsymbol{\xi}} p(\boldsymbol{\xi}) \ev{\op{A}}{\psi(\boldsymbol{\xi})}.
\end{align}
Using this in Eq.~\eqref{ot-tr-eq}, we can write $\ev{O(t)}$ in terms of an integral over $\boldsymbol{\xi}$ as
\begin{align}\label{sse-tr-sample-eq}
	\ev{O(t)} = \ev{\int \dd{\boldsymbol{\xi}} p(\boldsymbol{\xi}) \ev{\op{O}}{\Psi_{\sing,\boldsymbol{\xi}}(t)}},
\end{align}
where $\ket{\Psi_{\sing,\boldsymbol{\xi}}(0)} = \ket{\sing}\otimes\ket{\psi(\boldsymbol{\xi})}$, and the state $\ket{\Psi_{\sing,\boldsymbol{\xi}}(t)}$ obeys the SSE, 
\begin{align}
	\dv{t} \ket{\Psi_{\sing,\boldsymbol{\xi}}(t)} = \left(-\frac{i}{\hbar}\op{H}(\sX(t)) -  \op{K}\right)\ket{\Psi_{\sing,\boldsymbol{\xi}}(t)}.
\end{align}
In practice, the integral over $\boldsymbol{\xi}$ is evaluated using Monte Carlo sampling, which is seamlessly combined with the sampling of the stochastic variables $\sX(t)$. These equations define the SSE method with trace sampling.

As the number of hyperfine-coupled nuclear spins increases, the combined electronic and nuclear spin Hilbert space dimensionality of the radical pair, $D=4Z$, increases exponentially. Because the density operator $\op{\rho}(t)$ has $D^2$ matrix elements, the computational effort of directly solving Eq.~\eqref{haber-eq}, or the SLE derived from it, grows prohibitively large for models of radical pairs containing a realistic number of nuclear spins. By employing the SSE, the computational effort can be significantly reduced, especially when stochastic trace sampling is combined with the inevitable stochastic sampling of the fluctuations in the variables $\sX(t)$. Overall, the scaling of the SSE method is $\mathcal{O}(N_t M D\log D)$, where $N_t$ is the number of evolution time steps and $M$ is the number of Monte Carlo samples. For comparison, the SLE method requires $\mathcal{O}(N_t N_b^2 D^2 \log D)$ operations, where $N_b$ is the number of basis functions in $\sX$ needed in the expansion of $\op{\rho}(t,\sX)$. The SSE method is therefore faster by a factor of $\mathcal{O}(N_b^2D/M)$, which can become very significant indeed for radical pairs with many nuclear spins (exponentially large $D$).

The efficiency of the trace sampling depends on the choice of nuclear spin states $\ket{\psi(\boldsymbol{\xi})}$ used to resolve the identity operator in Eq.~\eqref{gen-res-id-eq}, which we have thus far not specified. We shall now present two methods for sampling, one based on the spin coherent states we have used before,\cite{Lewis2016,Fay2017,Lindoy2018,Fay2020,Lindoy2020,Keens2020} and the other based on $SU(Z)$ coherent states. Ideally, the choice of $\ket{\psi(\boldsymbol{\xi})}$ will yield a sampling method that is self-averaging, meaning that the statistical error in the sampled trace is $\mathcal{O}(1/\sqrt{M Z})$ and therefore \textit{exponentially} convergent in the number of nuclear spins. In the appendix, we derive a criterion for a sampling method to be self-averaging, and show that the $SU(Z)$ coherent state method satisfies this criterion. More generally, we expect that almost any trace sampling method will become self-averaging for observables evaluated after a sufficient period of radical pair spin dynamics, for reasons explained in the appendix. 

\subsubsection{Spin coherent states}

The spin coherent states $\ket{\Omega}\equiv\ket{\theta,\phi}$, for a nuclear spin with total angular momentum quantum number $I$, are the rotations of the $\ket{I,M_I = +I}$ $z$ projection state such that the new quantisation axis lies along $\vb{n}(\Omega) = (\sin\theta\cos\phi,\sin\theta\sin\phi,\cos\theta)$,
\begin{align}
	\ket*{\Omega^{(I)}} = \cos(\theta/2)^{2I} \exp(\tan(\theta/2)e^{i\phi}\op{I}_-) \ket{I,I}.
\end{align}
The identity operator on the Hilbert space of the nuclear spin can be resolved in terms of an integral over these states as
\begin{align}
	\op{1} &= \frac{2I+1}{4\pi}\int_0^{2\pi}\dd{\phi} \int_0^\pi \dd{\theta}\sin\theta \dyad*{\Omega^{(I)}} \label{coh-res-id-eq} \\
	&= (2I+1) \int\dd{\Omega}p(\Omega)\dyad*{\Omega^{(I)}}, 
\end{align}
where $p(\Omega) = \sin(\theta)/4\pi$ and $\int\dd{\Omega}= \int_0^{2\pi}\!\!\dd{\phi} \int_0^\pi\!\! \dd{\theta}$. In view of this, the full nuclear spin identity operator of a radical pair containing multiple nuclear spins can be resolved as 
\begin{align}
	\op{1} = Z \int \dd{\boldsymbol{\Omega}} p(\boldsymbol{\Omega})\dyad{\boldsymbol{\Omega}},
\end{align}
where $\ket{\boldsymbol{\Omega}}$ is a spin coherent state product, $\ket{\boldsymbol{\Omega}} = \bigotimes_{1=1}^{2}\bigotimes_{k=1}^{N_i}\ket*{\Omega^{(I_{ik})}_{ik}}$, and the integral is over the probability distribution for each spin coherent state $\int\dd{\boldsymbol{\Omega}} p(\boldsymbol{\Omega})= \prod_{i=1}^2\prod_{k=1}^{N_i}\int\dd{\Omega_{ik}}p(\Omega_{ik})$. The coherent state products $\ket{\boldsymbol{\Omega}}$ can therefore be used to sample the trace over nuclear spin states as in Eq.~(10).

\subsubsection{$SU(Z)$ coherent states}

An alternative trace sampling method is to use $SU(Z)$ coherent states ($SU(N)$ coherent states with $N=Z$),\cite{Nemoto2000} which we denote $\ket{\vb{Z}}$. $\vb{Z}$ is a vector of $Z$ complex numbers $Z_n = X_n+ iY_n$ and the $\ket{\vb{Z}}$ state is defined in a chosen basis as 
\begin{align}
	\ket{\vb{Z}} = \sum_{n=1}^Z \ket{n}Z_n,
\end{align}
such that $\braket{\vb{Z}}=1$. The identity operator on the nuclear spins can then be resolved as 
\begin{align}
	\op{1} = Z \int_{\mathbb{R}^Z} \dd{\vb{X}}\int_{\mathbb{R}^Z} \dd{\vb{Y}}\frac{\delta(|\vb{Z}|-1)}{\mathcal{S}_{2Z}}\dyad{\vb{Z}},
\end{align}
where $\mathcal{S}_{2Z}$ is the surface area of a $2Z$ dimensional hypersphere of unit radius. It follows that we can also sample the nuclear spin state trace as in Eq.~(10) by sampling $SU(Z)$ coherent states from the distribution $p(\vb{Z})=\delta(|\vb{Z}|-1)/\mathcal{S}_{2Z}$. Because this distribution is invariant under unitary transformations of the vector $\vb{Z}$, $\vb{Z} \to \vb{U}\vb{Z}$, this sampling method is independent of the choice of basis $\ket{n}$ in the definition of $\ket{\vb{Z}}$. This property is very important in proving that $SU(Z)$ coherent state sampling is self-averaging, as we show in the appendix. (In practice, one can sample $\vb{Z}$ from the distribution $p(\vb{Z})$ simply by sampling $2Z$ independent normal deviates $X_n$ and $Y_n$ and normalising the resulting $\vb{Z}$ vector.)

\subsection{Approximate methods}

In Sec.~III we shall compare the results obtained using the (formally exact) SSE method described above with those obtained using more approximate treatments of the spin dynamics and spin relaxation in radical pairs, which we shall now briefly summarise for completeness.

As a starting point for introducing these approximate treatments, it is convenient to divide the spin Hamiltonian into a time-independent reference part $\op{H}_0 = \ev{\op{H}(t)}$ and a time-dependent fluctuation $\op{V}(t) = \op{H}(t)-\ev{\op{H}(t)}$. We shall assume that the latter can be treated  perturbatively and can be expressed as a sum of fluctuating terms,
\begin{align}
	\op{V}(t) = \sum_{j} f_j(t) \op{A}_j,
\end{align}
in which $f_j(t) \equiv f_j(\sX(t))$ is a scalar function of $\sX(t)$ and $\op{A}_j$ is some operator on the radical pair spin states. The correlation functions of the fluctuating terms are given by $g_{jk}(t) = \ev{f_j(t)^*f_k(0)}$, and the Fourier-Laplace transforms of these are denoted $J_{jk}(\omega) = \int_0^\infty \dd{t} g_{jk}(t)e^{i\omega t}$.

\subsubsection{Nakajima-Zwanzig theory}

Second order Markovian Nakajima-Zwanzig (NZ) theory\cite{Nakajima1958,Zwanzig1960} gives the following master equation for the ensemble-averaged density operator $\ev{\op{\rho}(t)}$,
\begin{align}
	\dv{t}\ev{\op{\rho}(t)} = \pL_0 \ev{\op{\rho}(t)} + {\pR}_\mathrm{NZ}\ev{\op{\rho}(t)},
\end{align}
where $\pL_0 = -(i/\hbar)[\op{H}_0,\cdot ]-\{\op{K},\cdot\}$ and 
\begin{align}\label{nz-rel-op-eq}
	{\pR}_\mathrm{NZ} = -\sum_{jk}\int_0^\infty \dd{\tau} g_{jk}(\tau)\pazocal{A}_j^\dag e^{\pL_0\tau} \pazocal{A}_k,
\end{align}
with $\pazocal{A}_k = -(i/\hbar)[\op{A}_k,\cdot]$ and $\pazocal{A}_j^\dag = (i/\hbar)[\op{A}_j^\dag,\cdot]$.\cite{Fay2019a} The treatment of relaxation in this master equation is closely related to that in the more commonly used Bloch-Redfield-Wangsness relaxation theory,\cite{Wangsness1953,Redfield1965,Goldman2001} but it alleviates the severe positivity problem of that theory in the static disorder (slow nuclear motion) limit.\cite{Fay2019a} 

\subsubsection{The Lindblad equation}

By making the extreme-narrowing approximation to Eq.~\eqref{nz-rel-op-eq}, in which we assume that $g_{jk}(\tau)$ decays to zero on a time-scale much faster than the dynamics generated by $\pL_0$,\cite{Kattnig2016} we obtain a Lindblad (LB) type master equation\cite{Breuer2007} for $\ev{\op{\rho}(t)}$,
\begin{align}
	\dv{t}\ev{\op{\rho}(t)} &= \pL_0 \ev{\op{\rho}(t)} + \pR_\mathrm{LB}\ev{\op{\rho}(t)},
\end{align}
in which
\begin{align}
	\pR_\mathrm{LB} = \sum_{jk} &\gamma_{jk} \left(\op{A}_k \ev{\op{\rho}(t)} \op{A}_j^\dag - \frac{1}{2}\left\{\op{A}_j^\dag\op{A}_k,\ev{\op{\rho}(t)}\right\}\right),
\end{align}
with $\gamma_{jk} = 2 J_{jk}(0)/\hbar^2$. Although we have derived it here as an approximation to the perturbative Nakajima-Zwanzig equation in Eq.~(20), this form of quantum master equation exactly preserves positivity of the ensemble-averaged density operator,\cite{Breuer2007} and it is commonly used to model relaxation effects in radical pairs.\cite{Kattnig2016,Lukzen2017,Steiner2018,Player2020}

\subsubsection{Schulten-Wolynes theory}

Schulten-Wolynes (SW) theory\cite{Schulten1978} is a semiclassical approximation that circumvents the exponential scaling of quantum mechanics by replacing the quantum mechanical nuclear spin operators with classical vectors, $\op{\vb{I}}_{ik} \to \vb{I}_{ik}$. These vectors are taken to have the semiclassical  lengths $\sqrt{I_{ik}(I_{ik}+1)}$ and are each sampled uniformly from the surface of a sphere. This approximation generally becomes more accurate as the number of coupled nuclear spins increases. We have previously combined the SW approximation with the NZ theory of relaxation in a rigorously consistent way to model the spin dynamics of radical pairs.\cite{Fay2019a} 

\section{Results}\label{results-sec}

Here we demonstrate the utility of the SSE applied to radical pair spin dynamics by considering two sets of model problems. The first is based on a recent study by Keens \& Kattnig into relaxation effects on \ce{FAD^{$\bullet-$}-Z^{$\bullet$}} and \ce{FAD^{$\bullet-$}-W^{$\bullet+$}} radical pairs,\cite{Keens2020} and the second is based on our own recent study\cite{Fay2019a} of the \ce{DMJ^{$\bullet+$}-An-Ph_$\text{n}$-NDI^{$\bullet-$}} radical pairs investigated experimentally by Scott \textit{et al.}\cite{Scott2009}

\subsection{\ce{FAD^{$\bullet-$}-X^{$\bullet$}} radical pairs}\label{results-fadx-sec}
\begin{figure}
	\includegraphics[width=0.47\textwidth]{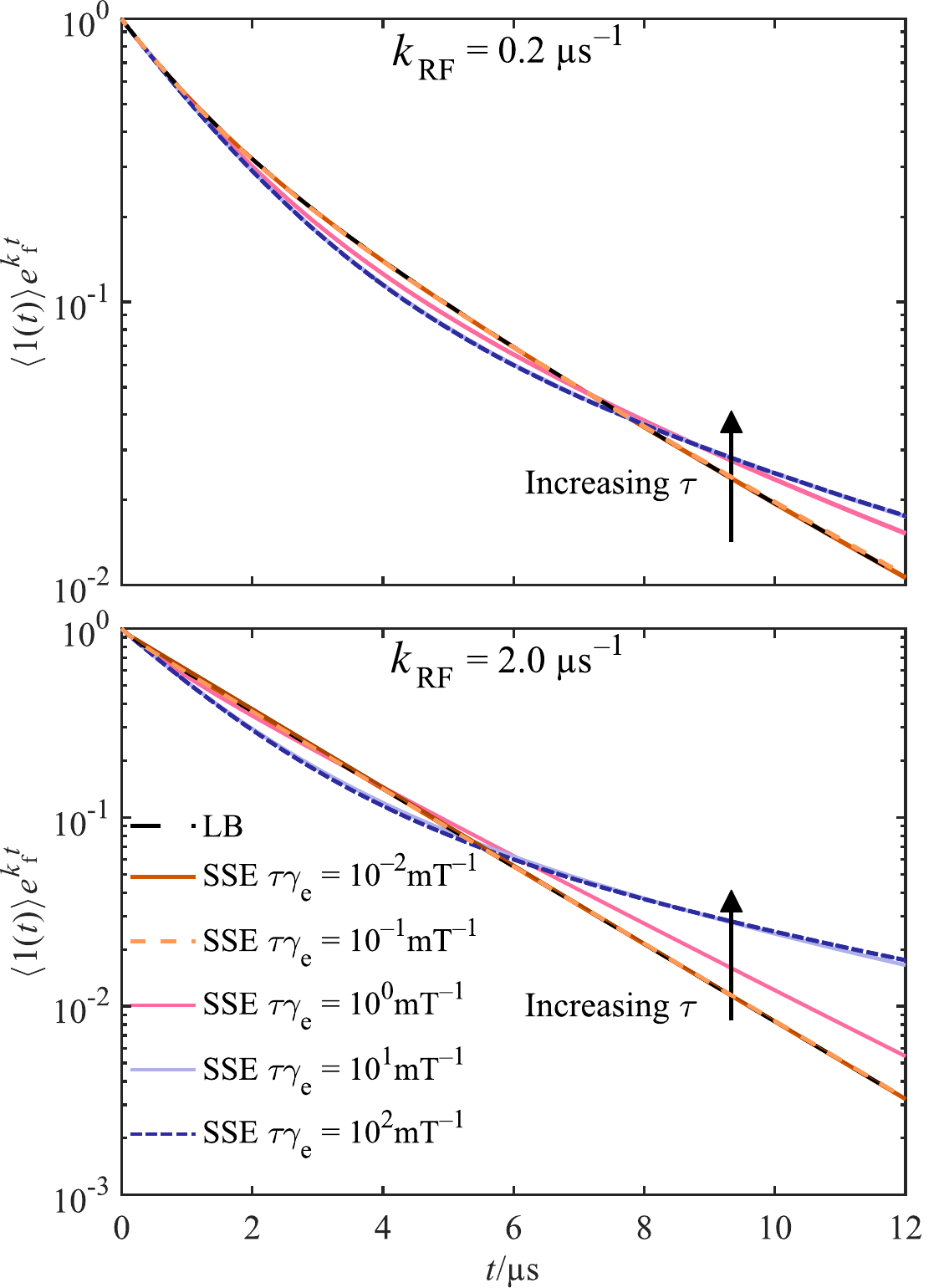}
	\caption{Rescaled survival probabilities of \ce{FAD^{$\bullet-$}-W^{$\bullet+$}} radical pairs with a total of 8 coupled nuclear spins (four in each radical), calculated with the Lindblad (LB) equation and the  SSE with a range of fluctuation timescales. In the top panel $k_{\mathrm{RF}} = 0.2 \ \muup\mathrm{s}^{-1}$, and in the bottom panel $k_{\mathrm{RF}} =  2.0 \ \muup\mathrm{s}^{-1}$.  $M=1024$ Monte Carlo samples were used in each simulation, giving error bars (2 standard errors in the mean) narrower than the widths of the plotted lines.}\label{fadw-fig}
\end{figure}

As a first example application of the SSE with coherent state sampling, we consider a set of \ce{FAD^{$\bullet-$}-X^{$\bullet$}} radical pairs with random fields relaxation. Here \ce{X^{$\bullet$}} is either \ce{W^{$\bullet+$}}, a tryptophan radical, or \ce{Z^{$\bullet$}}, a radical with no hyperfine coupled nuclear spins, both of which are commonly examined model radical pairs in the context of avian magnetoreception. As a test-bed for their jump trajectory Monte Carlo wavefunction method for finding solutions the Lindblad equation, Keens \& Kattnig used these radical pairs with a random fields (RF) model of spin relaxation.\cite{Keens2020} 

The stochastically evolving variables $\sX(t)$ in this case are six fluctuating magnetic field components $\Delta B_{i\alpha}(t)$ (three for each electron spin), such that the randomly fluctuating term in the spin Hamiltonian is 
\begin{align}
	\op{V}(t) = \sum_{i=1,2}\sum_{\alpha = x,y,z} g_\el \mu_\mathrm{B} \Delta B_{i\alpha}(t) \op{S}_{i\alpha},
\end{align}
where $\ev{\Delta B_{i\alpha}(t)\Delta B_{j\beta}(0)} = \delta_{ij}\delta_{\alpha\beta}\ev{\Delta B^2} g(t)$. In the extreme narrowing limit this fluctuation term gives a Lindblad relaxation superoperator of the form
\begin{align}
	\pR_\mathrm{RF} \!\ev{\op{\rho}(t)}\!= \!k_{\mathrm{RF}}\!\!\sum_{i=1,2}\sum_{\alpha = x,y,z} \!\!\left(\op{S}_{i\alpha}\!\ev{\op{\rho}(t)}\!\op{S}_{i\alpha} \!-\!\frac{1}{4}\ev{\op{\rho}(t)}\right)\!,
\end{align}
where the relaxation rate is $k_{\mathrm{RF}} = 2 \gamma_\el^2 \ev*{\Delta B^2} \tau$, in which $\gamma_\el$ is the gyromagnetic ratio of a free electron and the fluctuation timescale is $\tau = \int_0^\infty g(t)\dd{t}$. 

From the perspective of treating relaxation with the Lindblad equation, the precise details of the random field fluctuations are irrelevant, but beyond the extreme-narrowing limit they become important. In our SSE calculations, we choose the random fields to obey an overdamped Langevin equation of the form
\begin{align}
	\dv{t} \Delta B_{i\alpha}(t) = -\frac{1}{\tau} \Delta B_{i\alpha}(t) + \xi_{i\alpha}(t),
\end{align}
where $\xi_{i\alpha}(t)$ are independent delta-correlated stationary Gaussian processes, obeying $\ev{\xi_{i\alpha}(t)} = 0$ and $\ev{\xi_{i\alpha}(t)\xi_{i\alpha}(t')} = (2\ev*{\Delta B^2}/\tau) \delta(t - t')$. With this choice, $g(t) = e^{-t/\tau}$. The initial values of $\Delta B_{i\alpha}(t)$ at $t=0$ are sampled from the stationary distribution of this overdamped Langevin equation, $p(\Delta B_{i\alpha}) \propto \exp[-\Delta B_{i\alpha}^2/(2\ev*{\Delta B ^2})]$,\cite{Breuer2007} and the subsequent values of $\Delta B_{i\alpha}(t)$ that are used in the SSE simulations are obtained by numerical integration of Eq.~(26) as described in Appendix~A.

As a first example, we consider a \ce{FAD^{$\bullet-$}-W^{$\bullet+$}} radical pair, with four hyperfine coupled nuclei in each radical, with a static magnetic field of strength 1 mT aligned along the positive $z$ axis. The hyperfine coupling tensors are taken from Ref.~\onlinecite{Keens2020} and are given in the Supplementary Material. The rate constants in the model are $k_\sing = k_\mathrm{b} + k_\mathrm{f}$ and $k_\trip = k_\mathrm{f}$, so the uniform exponential decay due to $k_\mathrm{f}$ can be factored out, and $k_\mathrm{b}$ is set to $k_\mathrm{b} = 2.0\ \muup\mathrm{s}^{-1}$. We consider two examples where the extreme-narrowing limit relaxation rates are $\gamma_\mathrm{RF} = 0.2\ \muup\mathrm{s}^{-1}$ and $\gamma_\mathrm{RF} = 2.0\ \muup\mathrm{s}^{-1}$, but the time scale $\tau$ is varied, and $\ev*{\Delta B^2}$ is fixed at  $\ev*{\Delta B^2} = k_{\mathrm{RF}}/(2\gamma_\el^2\tau)$. 

In Fig.~\ref{fadw-fig}, the total surivival probability of the radical pair with the symmetric part of the decay factored out, $\ev{1(t)}e^{k_\mathrm{f}t}$ (the same observable considered in Ref.~\onlinecite{Keens2020}), is shown as a function of time, for a range of values of $\tau\gamma_\el$, with the extreme-narrowing limit Lindblad results also shown for comparison. All the spin coupling parameters are on the order of 1 mT, so as $\tau \gamma_\el$ is varied between $10^{-2}$ to $10^2\ \mathrm{mT}^{-1}$, there is a transition in relaxation behaviour from the extreme-narrowing limit to the static disorder limit. We see that the LB equation is almost quantitatively accurate when compared to the SSE when $\tau \gamma_\el$ is between $10^{-2}\ \mathrm{mT}^{-1}$ and $10^{-1}$ mT, but for $\tau \gamma_\el\geq10^0\ \mathrm{mT}^{-1}$ significant deviations from the extreme-narrowing limit LB equation can be seen in both the short time and long time decay of $\ev{1(t)}e^{k_\mathrm{f}t}$. The short time decay rate increases as $\tau$ is increased, whereas the long time decay rate decreases. This significantly alters the long time survival probability. The effect becomes more pronounced as $k_{\mathrm{RF}}$ increases and relaxation makes a larger contribution to the interconversion between singlet and triplet states, which react at different rates. 

Not only does the SSE capture relaxation effects that cannot be captured by the Lindblad equation: it also provides a more efficient way to do the calculation. The computational effort of the SSE calculation is $\mathcal{O}(N_t M D\log D)$, whereas solving the Lindblad equation has an effort of $\mathcal{O}(N_t D^2 \log D)$. Since the number of Monte Carlo samples needed for convergence, $M$, is typically less than the Hilbert space dimensionality, $D$, the SSE approach is more efficient. Furthermore, the SSE has at best an $\mathcal{O}(D)$ memory requirement, whereas solving the Lindblad equation has an $\mathcal{O}(D^2)$ memory requirement, and this is often the limiting factor in calculations on large spin systems. The Monte Carlo wavefunction method of Keens \& Kattnig\cite{Keens2020} reduces the computational effort of the Lindblad calculation to $\mathcal{O}(N_t M' D \log D)$, and the memory requirement to $\mathcal{O}(D)$. However, the number of Monte Carlo samples needed to converge this method ($M'$) is typically far larger than the number needed to converge the SSE ($M$). For example, in the examples examined in Ref.~\onlinecite{Keens2020}, at least $M'=16,000$ Monte Carlo samples were used, and as many as 476,800. In our tests of the SSE with coherent state sampling, we have found that far fewer Monte Carlo samples can be used to obtain well converged results.

To illustrate this, we consider a \ce{FAD^{$\bullet-$}-Z^{$\bullet$}} radical pair with 12 coupled nuclear spins in the \ce{FAD^{$\bullet -$}} radical, with $\tau\gamma_\el = 1 \ \mathrm{mT}^{-1}$. Since this value of $\tau$ is in the intermediate regime, relaxation effects in this example cannot be captured with an extreme-narrowing limit Lindblad equation (or the Monte Carlo wavefunction method), and with a Hilbert space dimension of $D = 36,864$, this problem cannot be treated with the SLE. In line with what was studied by Keens \& Kattnig,\cite{Keens2020} we consider the rescaled survival probability $\ev{1(t)}e^{k_\mathrm{f}t}$, and we set $k_{\mathrm{RF}} = 0.2\ \muup\mathrm{s}^{-1}$ and $k_\mathrm{b} = 2.0 \ \muup\mathrm{s}^{-1}$.

\begin{figure}
	\includegraphics[width=0.5\textwidth]{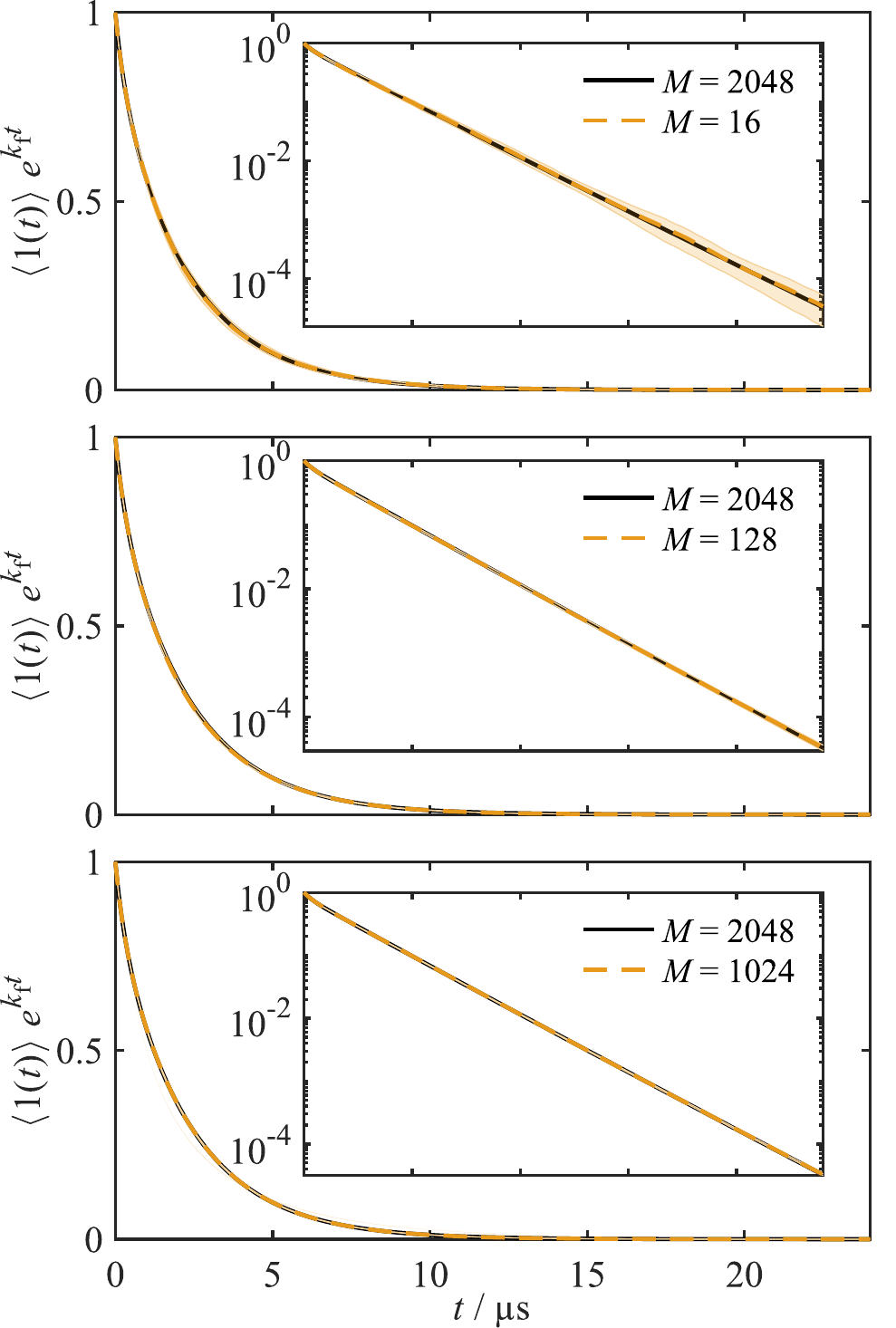}
	\caption{Rescaled survival probabilities of \ce{FAD^{$\bullet-$}-Z^{$\bullet$}} radical pairs with a total of 12 coupled nuclear spins, calculated with the SSE with different numbers of Monte Carlo samples $M$. Each curve represents the results from an \textit{independent} set of Monte Carlo samples. The shaded error bars around each of the SSE curves are $\pm 2$ standard errors in the mean at each time. The inserts show the same data over the same time scale with a log scale on the $y$ axis.}\label{fadz-fig}
\end{figure}

The results in Fig.~\ref{fadz-fig} show the rescaled survival probability as a function of time, for $M=16$, 128, 1024 and 2048. Each curve corresponds to an independent set of Monte Carlo samples, and the shaded areas indicate 2 standard errors in the mean. These results show that reasonably well converged calculations can be performed with as few as 16 Monte Carlo samples, and that the results are converged to graphical accuracy on a log scale with 1024 samples. This implies that the SSE with coherent state sampling can be used to perform comparable calculations several orders of magnitude more efficiently than the jump quantum trajectory method of Keens \& Kattnig.\cite{Keens2020} 

This remarkable performance of the SSE compared to the jump trajectory method can be explained by noting two things. Firstly, the trajectories in the SSE method are continuous, whereas those in the jump trajectory method are discontinuous, which hinders convergence. The probability of a trajectory surviving to time $t$ in the jump trajectory method is $p(t)\approx e^{-\bar{k}t}$, where $\bar{k}$ is some average decay constant for the radical pair spin dynamics. Since only an exponentially small fraction of the sampled trajectories survive to longer times, the convergence is worse for the jump trajectory method at large $t$. In the SSE method, all trajectories contribute at all times, so this is not an issue. Secondly, there is a large degree of self-averaging for large spin systems in the SSE results,\cite{Weisse2006} a feature which is expanded on in the appendix. Furthermore, each SSE trajectory in the example we have considered in Fig.~2 samples the stochastic fluctuations in $\sX(t)$ over a time scale of $4000 \tau$, where $\tau$ is the correlation time of $\Delta B_{i\alpha}(t)$. The stochastic fluctuations in the Hamiltonian are therefore well sampled, and this does not limit the convergence of the SSE results.

\subsection{\ce{DMJ^{$\bullet+$}-An-Ph_$\text{n}$-NDI^{$\bullet-$}} radical pairs}\label{results-dmjndi-sec}
\begin{figure*}[ht]
	\begin{minipage}[c]{0.74\textwidth}
		\centering
		\includegraphics[width=\textwidth]{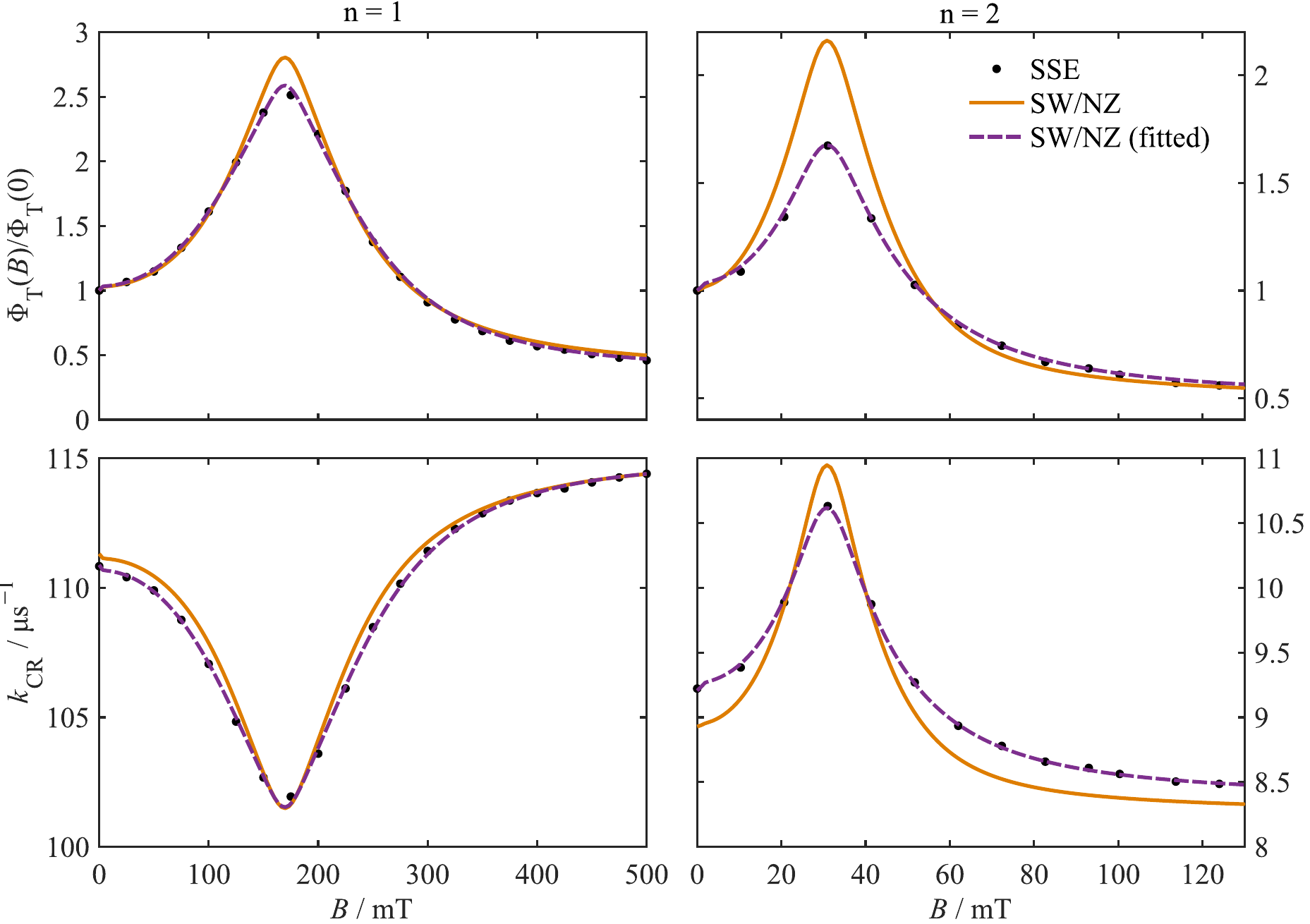}
	\end{minipage}
	\hfill
	\begin{minipage}[c]{0.25\textwidth}
		\caption{Relative triplet yields (top panels) and charge recombination rates (bottom panels) for the n=1 (left panels) and n=2 (right panels) \ce{DMJ^{$\bullet+$}-An-Ph_$\text{n}$-NDI^{$\bullet-$}} radical pair models as a function of applied magnetic field strength $B$. Black dots are the SSE results, orange lines are the SW/NZ method results and the purple dashed lines are the re-fitted SW/NZ results with $k_\sing$, $k_\trip$, $\sigma_J$ and $\tau_J$ treated as free parameters.}\label{dmj-ndi-fig}
	\end{minipage}
\end{figure*}

We have recently presented a study of relaxation effects in {\ce{DMJ^{$\bullet+$}-An-Ph_$\text{n}$-NDI^{$\bullet-$}}} radical pairs in which we applied an approximate spin dynamics method based on a combination of the Schulten-Wolynes (SW) approximation to the hyperfine interactions and a perturbative Nakajima-Zwanzig (NZ) treatment of relaxation effects.\cite{Fay2019a} This involved using the SW/NZ method to fit various models of the relaxation to experimental magnetic field effect data on the relative triplet product yields and radical pair decay rates. Based on these fits, we shall now construct a simplified model of the n=1 and n=2 radical pairs for which we can simulate the relative triplet yields and radical pair decay rates using the SSE. We shall then use these simulations test the accuracy of the SW/NZ approximation, and also use the SSE data as ``simulated experimental data'' with which to test the accuracy fitting parameters using the SW/NZ method, as we did with the experimental data in our previous study.\cite{Fay2019a} 

The model we shall consider includes relaxation from rotational diffusion of the radical pair molecule, which modulates the anisotropic components of the hyperfine and $g$ tensors in both radicals, and relaxation due to modulation of the scalar coupling between the electron spins. The molecule is treated as a rigid body undergoing anisotropic rotational Brownian motion, so the Euler angles $\Upomega(t)$ describing the orientation of the molecule fluctuate with time. In this treatment, the coupling tensors $\vb{C}$ ($\vb{A}_{ik}$, $\vb{D}$, $\vb{g}_i$ etc.) fluctuate according to 
\begin{align}
	\vb{C}(\Upomega(t)) = \vb{R}(\Upomega(t))\cdot \vb{C}_\mathrm{mol} \cdot \vb{R}(\Upomega(t))^{-1},
\end{align}
in which $\vb{R}(\Upomega(t))$ is a rotation matrix defined such that vector operators in the molecular frame $\op{\vb{J}}_\mathrm{mol}$ are related to vector operators in the laboratory frame by $\op{\vb{J}}_\mathrm{lab}=\vb{R}(\Upomega(t))\op{\vb{J}}_\mathrm{mol}$, and $\vb{C}_\mathrm{mol}$ is the coupling tensor in the molecular frame. The scalar coupling modulation is modelled with a symmetric two site model, where $J = \ev{J} \pm \sigma_J $ in the two sites, and the rate of exchange between the two sites is $k_\mathrm{ex} = 1/(2\tau_J)$. The stochastic variables $\sX(t)$ are thus the Euler angles $\Upomega(t)$ and the current site the molecule $\mathsf{s}(t)$, which determines $J$. The model includes 15 coupled nuclear spins, plus the two electron spins, and as such it is beyond direct treatment with the SLE. Other details of the model parameters can be found in the Supplementary Material. 

The results of these simulations are shown Fig.~\ref{dmj-ndi-fig} for the relative triplet yields, $\Phi_\trip(B)/\Phi_\trip(0)$, and the radical pair recombination rate, $k_\mathrm{CR}$, which are defined in terms of ensemble averaged observables as 
\begin{align}
	\Phi_\trip &= k_\trip \int_0^\infty \ev{P_\trip(t)}\dd{t} , \\
	\frac{1}{k_\mathrm{CR}} &= \int_0^\infty  \ev{1(t)} \dd{t}.
\end{align}
The full set of model parameters used in the simulations are given as supplementary material. The relative triplet yields and radical pair recombination rates are the same observables as were reported in the experimental study by Scott \textit{et al.}.\cite{Scott2009}

The n=1 model has a lifetime about 10 times shorter, and an exchange splitting $2\ev{J}$ about 5 times larger, than the n=2 model. The electron spin coupling parameters are therefore smaller relative to the hyperfine interactions in the n=2 model, which presents a greater challenge for the SW/NZ method. This is borne out in the comparison of the SW/NZ and SSE simulations in Fig.~\ref{dmj-ndi-fig}. The results of the two simulations agree almost quantitatively across all applied fields for n=1, but not for n=2. The largest deviations for the n=1 model are at $B=0$ and at the $B=2\ev{J}/g_\el \mu_\mathrm{B}$ resonance, where they are are still less than about 10\%. Predictably, the quantitative deviations are larger in the n=2 case, especially on resonance in the relative triplet yield data. This can be understood by considering the error in the SW/NZ simulation of $k_\mathrm{CR}$ for the n=2 model. At $B=0$, $k_\mathrm{CR}$ is underestimated, and at $B=2\ev{J}/g_\el\mu_\mathrm{B}$ it is overestimated. Because $k_\mathrm{CR}^{-1} = (k_\trip^{-1}-k_\sing^{-1}) \Phi_\trip+k_\sing^{-1}$, these two errors are compounded in the  $B=2\ev{J}/g_\el\mu_\mathrm{B}$ resonance in the relative triplet yield.

We have also performed parameter fitting to the SSE data with the SW/NZ method to analyse the accuracy of the fitted parameters obtained with this method. The free parameters in the model are $k_\sing$, $k_\trip$, $\sigma_J$ and $\tau_J$, analogous to the parameter fitting we performed in Ref.~\onlinecite{Fay2019a} using the experimental data from Ref.~\onlinecite{Scott2009}, and these parameters were fitted by minimising the normalised mean square error as described previously.\cite{Fay2019a} The fitted SW/NZ results are also shown in Fig.~\ref{dmj-ndi-fig}. We see that the fitted results agree quantitatively with the SSE simulation data for both models (n=1 and n=2), which shows that the SW/NZ method does not miss any qualitative features in the exact SSE data. The original and fitted parameters are summarised in Table~\ref{dmj-ndi-params-tab}. The errors in the fitted rate constants $k_\sing$ and $k_\trip$ are very small (less than 6\%), with the errors being larger in the n=2 case where the SW/NZ method is less reliable (see above). The errors in the fitted $\sigma_J$ and $\tau_J$ parameters are larger, with the largest errors again observed in the n=2 case. This is probably due to the fact that $\tau_J$ is much shorter than the time scale of the spin dynamics, so the $2J$ fluctuations in this model are in the extreme narrowing limit. In this limit the $2J$ fluctuations cause singlet-triplet dephasing at a rate 
$k_{\sing\trip\mathrm{D}} = \left({2\sigma_J}/{\hbar}\right)^2 \tau_J$,
and therefore the observed magnetic field effects are only really dependent on one parameter, $\sigma_J^2 \tau_J$. The error in this parameter is $6\%$ for the n=1 model and $36\%$ for the n=2 model.  

\begin{table}[t]
	\centering
	\begin{tabular}{lcccc}
		\toprule
		& $k_\sing/\mathrm{ns}^{-1}$ & $k_\trip/\mathrm{ns}^{-1}$ &$(2\sigma_J / g_\el\mu_\mathrm{B})/\mathrm{mT}$ & $\tau_J/\mathrm{ns}$ \\
		\toprule 
		n=1 &  0.118 &  0.0301  & 201 & 0.00771\\
		n=1 (fitted) &  0.118 &  0.0293 & 146 & 0.0157 \\
		\midrule
		n=2 & 0.00770 & 0.0147 & 9.32 & 0.179 \\ 
		n=2 (fitted) & 0.00769& 0.0139 & 20.0 & 0.0532 \\
		\bottomrule
	\end{tabular}
\caption{Parameters used in the n=1 and n=2 models of \ce{DMJ^{$\bullet+$}-An-Ph_$\text{n}$-NDI^{$\bullet-$}} radical pairs. The \lq\lq fitted" parameters are the parameters obtained by using the SW/NZ method to fit the SSE simulation data.}\label{dmj-ndi-params-tab}
\end{table}

\section{Concluding remarks}\label{conc-sec}

In this paper, we have shown how to use the stochastic Schr\"odinger equation to model relaxation effects on radical pair recombination reactions. The method is exactly consistent with the full stochastic Liouville equation which is commonly used to model relaxation effects, but the present method can treat much larger spin systems through the use of efficient trace sampling. (Note that the trace sampling used here is closely related to the stochastic resolution of the identity methods that have recently been used to accelerate electronic structure calculations.\cite{Baer2013,Takeshita2017,Dou2019} Trace sampling is clearly a very powerful and general technique for speeding up quantum mechanical calculations.)

We have illustrated the applicability of the SSE to problems in spin chemistry with two examples. In the first, we examined the effect of random field fluctuations on relaxation in a set of \ce{FAD^{$\bullet-$}-X^{$\bullet$}} radical pairs, based on the recent work of Keens \& Kattnig.\cite{Keens2020} We used this example to demonstrate the convergence of our method, showing that well converged results could be obtained with as few as 128 samples of the initial nuclear spin Hilbert space (of dimension $Z=9,216$). In the second example, we used the SSE to model relaxation effects in realistic models of \ce{DMJ^{$\bullet+$}-An-Ph_$\text{n}$-NDI^{$\bullet-$}} radical pairs. Here we used the method to validate the results of a computationally inexpensive Schulten-Wolynes/Nakajima-Zwanzig approximation which we have previously employed\cite{Fay2019a} to interpret experiments on these radical pairs.\cite{Scott2009} We found that, while the SW/NZ approximation does not perfectly capture all magnetic field effects, using it to fit unknown model parameters does yield reasonably accurate results for both the resulting magnetic field effects and the values of the fitted parameters.

Based on our results for these model problems, we anticipate that the SSE has the potential to find widespread use in spin chemistry and related fields, including the study of spins in quantum dots.\cite{Camenzind2018,Lindoy2018} In addition to aiding the interpretation of magnetic field effect experiments on radical pairs, the SSE could be used to study the effect of radio-frequency noise on radical pair reactions of relevance to avian magnetoreception, where it would provide a more general alternative to the Floquet based approaches which have been employed previously.\cite{Hiscock2016,Hiscock2017} The present method could trivially be extended to study relaxation effects in radical triad systems,\cite{Keens2018,Sampson2019} and it could also in principle be combined with recent symmetrisation techniques and used to study relaxation effects in larger spin systems.\cite{Lindoy2018,Lindoy2020} 

In the interpretation of magnetic field effect experiments on radical pairs, it is often necessary to fit several parameters in the model which are not known \textit{a priori}, for example the recombination rate constants $k_\sing$ and $k_\trip$, but fitting many parameters with the SSE can be difficult. This is because the SSE still involves doing calculations in the full spin Hilbert space, and fitting many parameters typically involves performing hundreds of magnetic field effect simulations. To avoid this expense, cheaper semiclassical methods,\cite{Schulten1978,Manolopoulos2013,Lewis2014} perturbative treatments of spin relaxation,\cite{Fay2019a} or kinetic master equations\cite{Steiner2018,Fay2019,Mims2019,Riese2020} can be used to fit the model parameters, followed by a one-shot SSE calculation to verify their accuracy, as we have illustrated in Sec.~III.B. 

While the SSE clearly provides a powerful tool for studying relaxation effects in radical pair systems, there are effects that the current approach cannot capture. In particular, finite temperature effects cannot be described with the present method. Methods do exist which can model finite temperature effects in an approximate way based on modified versions of Redfield theory\cite{Goldman2001,Bengs2020} and the Stochastic Liouville equation,\cite{Vega1975} but neither of these methods can treat both non-perturbative, intermediate time scale relaxation effects and finite temperature effects accurately. Furthermore, it is not clear how to accurately and consistently treat finite temperature effects in a system consisting of radicals recombining asymmetrically (i.e., with $k_\sing \neq k_\trip$). It has been noted that finite temperature effects can play an important role in systems far from equilibrium,\cite{Bengs2020} as is the case for radical pairs generated by photo-excitation, so these effects may become particularly important for radical pair reactions at low temperatures. It remains an open question how to address this shortcoming of the SSE in general, although for some simple models of relaxation the hierarchical equations of motion (HEOM) approach can be used,\cite{Takahashi2020} and other perturbative SSE approaches have also been suggested.\cite{Biele2014} When finite temperature effects can be ignored, however, we firmly believe that the SSE, combined with efficient trace sampling, should be the method of choice for modelling spin relaxation effects in radical pairs with a significant number of hyperfine coupled nuclear spins.

\begin{acknowledgements}
We are grateful to Daniel Kattnig for providing useful information on the simulations performed in Ref.~\onlinecite{Keens2020}. Thomas Fay is supported by a Clarendon Scholarship from Oxford University, an E.A. Haigh Scholarship from Corpus Christi College, Oxford, and by the EPRSC Centre for Doctoral Training in Theory and Modelling in the Chemical Sciences, EPSRC Grant No. EP/L015722/1. Lachlan Lindoy was supported by a Perkin Research Studentship from Magdalen College, Oxford, an Eleanor Sophia Wood Postgraduate Research Travelling Scholarship from the University of Sydney, and by a James Fairfax Oxford Australia Scholarship. Both Thomas Fay and Lachlan Lindoy also acknowledge support from the Air Force Office of Scientific Research (Air Force Materiel Command, USAF award no. FA9550-14-1-0095).
\end{acknowledgements}

\section*{Supplementary Material}

The Supplementary Material includes plain text files of all parameters used in the simulations presented in this paper.

\section*{Data Availability}

The data that support the findings of this study are available in the paper and the supplementary material.

\appendix
\section{Integrating the Stochastic Schr\"odinger equation}

Here we outline the algorithms we have employed to integrate the SSE and provide the integration parameters we have used in our simulations.

\subsection{Integrating the spin state}

The evolution operator $\op{U}(t_1,t_0)$ which evolves the spin state under a fluctuating Hamiltonian from $t_0$ to $t_1$ is given by
\begin{align}
	\op{U}(t_1,t_0) = \mathsf{T}\exp[\int_{t_0}^{t_1} \dd{\tau}\left(-\frac{i}{\hbar}\op{H}(\sX(\tau)) - \op{K}\right)]. 
\end{align}
For small time differences $t_1-t_0 = \delta t$, this can be approximated by the lowest order Magnus expansion term\cite{Magnus1954}  
\begin{align}
	\op{U}(t_0+\delta t,t_0) &\approx \exp[\int_{t_0}^{t_0+\delta t} \dd{\tau}\left(-\frac{i}{\hbar}\op{H}(\sX(\tau)) - \op{K}\right)] \\
	&= \exp[-i\op{\Omega}(t_0+\delta t,t_0)\delta t].
\end{align}
Hence the state can be evolved from $t_0$ to $t_1$ by evolving it with the generator 
\begin{align}
	\begin{split}
	\op{\Omega}(t_0+\delta t,t_0) &= \frac{1}{\hbar}\op{H}_0 - i \op{K} \\
	&- \frac{1}{\hbar\delta t} \sum_{j}\op{A}_j\int_{t_0}^{t_0+\delta t} f_j(\sX(\tau))\dd{\tau}.
	\end{split}
\end{align}
This evolution can be performed using a variety of algorithms, and in our case we use a fourth order short-iterative Arnoldi algorithm.\cite{Pollard1994} This requires us to find $\int_{t_0}^{t_0+\delta t} f_j(\sX(\tau))\dd{\tau}$, which is just an integral of a function of $\sX(t)$ for a given realisation of the stochastic fluctuations. All that remains is thus to find a way to evolve the stochastic variables. 

\subsection{Integrating stochastic variables}

We have used different algorithms for different models of the stochastic fluctuations. In each case, the trapezoidal rule was used to numerically integrate $f_j(\sX(\tau))$ to obtain the short time generator $\op{\Omega}(t_0+\delta t,t_0)$. 

For the over-damped Langevin equation used to model random field fluctuations in the \ce{FAD^{$\bullet -$}-X^{$\bullet$}} radical pair, the short time evolution of the random field components from $t_0$ to $t_0 + \delta t$ can be approximated by
\begin{align}
	\begin{split}
	\Delta B_{i\alpha}(t_0 + \delta t) &\approx \frac{1-\gamma_\el^2\ev{\Delta B^2}\delta t /(2\tau^2)}{1+\gamma_\el^2\ev{\Delta B^2}\delta t /(2\tau^2)}\Delta B_{i\alpha}(t_0) \\
	&+ \frac{1}{1+\gamma_\el^2\ev{\Delta B^2}\delta t /(2\tau^2)}\zeta_{i\alpha}(t_0,\delta t),
	\end{split}
\end{align}
where $\zeta_{i\alpha}(t_0,\delta t)$ is a random variable sampled from a normal distribution with zero mean and variance $2 \ev{\Delta B}^2\delta t/\tau$. This integrator can be derived straightforwardly using the ideas in Ref.~\onlinecite{Gronbech-Jensen2013}. 

For the $\ce{DMJ^{$\bullet+$}-An-Ph_$\text{n}$-NDI^{$\bullet-$}}$ radical pair models, the rotational Brownian motion was evolved using the integrator described in Ref.~\onlinecite{Delong2015}. The hopping in the two-site model was treated as a discrete time Markov chain, with a time interval of $\delta t$ between time points and with a transition probability of $p_\mathrm{hop} = 1-\exp(-k_\mathrm{ex}\delta t)$ at each step of the Markov chain.

\subsection{Integration parameters}

In the \ce{FAD^{$\bullet -$}-X^{$\bullet$}} model calculations, we used a time step $\delta t = t_1-t_0$ of $0.5\ \mathrm{ns}$, and a time step 100 times shorter for the evolution of the fluctuating field variables $\Delta B_{i\alpha}(t)$. The Lindblad calculations were performed using an adaptive short iterative Arnoldi integrator,\cite{Pollard1994} with a Krylov subspace dimension of 32. The Krylov subspace was updated when the coefficient of the last Krylov vector reached $10^{-8}$ of the 2-norm of the Liouville vector. 

In the SSE calculations on the \ce{DMJ^{$\bullet+$}-An-Ph_$\text{n}$-NDI^{$\bullet-$}} radical pair models, we used a time step of $0.1\ \mathrm{ns}$ for the spin state evolution and a time step 100 times smaller for the evolution of the stochastic variables. The simulations were run until the total survival probability of the radical pair had decayed to less than $10^{-5}$, and 256 Monte Carlo samples were used for each applied field strength.

\section{Trace sampling}

In this section we shall discuss the efficiency of trace sampling, and attempt to explain the differences in sampling efficiency between different trace sampling schemes. This analysis is based primarily on material in Ref.~\onlinecite{Weisse2006}.

\subsection{Convergence properties}

With any trace sampling approach, we are attempting to approximately evaluate an expression of the form
\begin{align}
	\mu_A = \frac{1}{Z}\tr[\op{A}] = \frac{1}{Z}\sum_{n=1}^Z \ev{\op{A}}{n},
\end{align}
in which $\op{A}$ is a nuclear spin operator, $\tr[\cdots]$ denotes the nuclear spin partial trace, the set of $\ket{n}$ states form a complete basis for the nuclear spin Hilbert space, and $Z$ is the dimensionality of this space. In our case, $\op{A}$ is an operator of the form
\begin{align}\label{A-op-gen-eq}
	\op{A} = \int_0^\infty\ev{\tr_\el[\op{\sigma}_\el\op{U}(\tau)^\dag\op{O} \op{U}(\tau)]}f(\tau) \dd{\tau}
\end{align}
where, as in the main text, $\ev{\cdots}$ denotes the average over stochastic fluctuations, $\op{U}(\tau)$ is the propagator, $\op{\sigma}_\el$ is the electron spin density operator, $\op{O}$ is an observable operator, $\tr_\el[\cdots]$ denotes the partial trace over the electron spins, and $f(\tau)$ is an arbitrary function of $\tau$. For example, for the $\ev{1(t)}$ observable considered in our \ce{FAD^{$\bullet -$}-X^{$\bullet$}} radical pair calculations, $\op{O} = \op{1}$, $\op{\sigma}_\el = \op{P}_\sing$ and $f(\tau) = \delta(t-\tau)$. For simplicity in the following analysis, we will assume that we have evaluated the average over stochastic variables $\ev{\cdots}$ exactly. In practice this is not the case, but for the purpose of analysing the efficiency of the trace sampling we can make this assumption.

In performing trace sampling, we approximate $\mu_A$ by an estimator $\Theta_A$. This is the result of an $M$ sample simulation, and is defined as
\begin{align}
	\mu_A \approx \Theta_A = \frac{1}{M}\sum_{r=1}^M \ev*{\op{A}}{\psi^{(r)}},
\end{align}
where the states $\ket*{\psi^{(r)}}$ are random normalised nuclear spin states. These states will in general be parametrised by some set of variables $\boldsymbol{\xi}^{(r)}$, i.e.$\ket*{\psi^{(r)}} = \ket*{\psi(\boldsymbol{\xi}^{(r)})}$, where each set of variables $\boldsymbol{\xi}^{(r)}$ is sampled randomly from an identical distribution. We will denote the full set of $M$ independent sets $\boldsymbol{\xi}^{(r)}$ of random variables as $\boldsymbol{\xi} = (\boldsymbol{\xi}^{(1)},\dots,\boldsymbol{\xi}^{(M)})$. For example, in the case of coherent state sampling, we generate a set of $M$ sets of random orientations of the nuclear spin vectors, so $\boldsymbol{\xi}^{(r)} = \boldsymbol{\Omega}^{(r)}$, where $\boldsymbol{\Omega}^{(r)}$ is the $r$th set of nuclear spin orientations generated in the simulation. Overall this means that $\Theta_A$ is itself a random variable, and we will be interested in the distribution of $\Theta_A$, in particular how its variance scales with $Z$. We will denote the average result of an $M$ sample simulation by $\ev{\dots}_M$. This corresponds to averaging over all possible values of the random variables in $\boldsymbol{\xi}$ such that
\begin{equation}
\ev{f(\boldsymbol{\xi})}_M = \int p(\boldsymbol{\xi}^{(1)})\,{\rm d}\boldsymbol{\xi}^{(1)}\cdots
\int p(\boldsymbol{\xi}^{(M)})\,{\rm d}\boldsymbol{\xi}^{(M)}\,\,f(\boldsymbol{\xi}),
\end{equation}
where $p(\boldsymbol{\xi}^{(r)})$ is the normalised probability density for $\boldsymbol{\xi}^{(r)}$ in Eq.~(9) (which is the same for all $r$). 

The coefficients of the randomly sampled states in the orthonormal basis $\ket{n}$, $c_n^{(r)} \equiv c_n(\boldsymbol{\xi}^{(r)})$, are 
\begin{align}
	c_{n}^{(r)} = \braket*{n}{\psi^{(r)}}.
\end{align}
When averaged over $\boldsymbol{\xi}$, we assume these coefficients obey the following relation
\begin{align}\label{c-corr-eq}
	\ev{{c_{n}^{(r)}}^* c_{n'}^{(r')}}_M = \frac{1}{Z}\delta_{r,r'}\delta_{n,n'},
\end{align}
where the factor of $1/Z$ naturally arises if the states are chosen to the normalised such that $\braket*{\psi^{(r)}}=1$. We note that if Eq.~\eqref{c-corr-eq} holds true in one basis, it must hold true in all other bases. From this equation, it is trivial to show that on average, the estimator for the trace $\Theta_A$ will be exactly the true quantum mechanical average,
\begin{align}
	\ev{\Theta_A}_M = \mu_A. 
\end{align}

In order to understand the convergence of the trace sampling, we should consider the fluctuations in the estimator, $\delta \Theta_A = \Theta_A - \mu_A$. Given that each set of random variables $\boldsymbol{\xi}^{(r)}$ is sampled independently and from the same distribution, the mean square fluctuation $\ev{\delta\Theta_A^2}_M$ can be evaluated as
\begin{align}
	\ev{\delta\Theta_A^2}_M &= \ev{\Theta_A^2}_M-\ev{\Theta_A}_M^2\nonumber\\ 
	\begin{split}
		&= \frac{1}{M^2}\sum_{r=1}^M\sum_{r'=1}^M \bigg( \ev{ {\ev*{\op{A}}{\psi^{(r)}}\!\!\ev*{\op{A}}{\psi^{(r')}}} }_M\\
		&\ \ \ - \ev{ {\ev*{\op{A}}{\psi^{(r)}}} }_M\ev{ { \ev*{\op{A}}{\psi^{(r')}} } }_M\bigg)
	\end{split}\nonumber\\
	\begin{split}
	 &= \frac{1}{M} \bigg( \ev{ {\ev*{\op{A}}{\psi^{(r)}}\!\!\ev*{\op{A}}{\psi^{(r)}}} }_M\\
	&\ \ \ - \ev{ {\ev*{\op{A}}{\psi^{(r)}}} }_M\ev{ { \ev*{\op{A}}{\psi^{(r)}} } }_M\bigg).
	\end{split}
\end{align}
Here the right-hand side is independent of $r$ because each set of variables ${\boldsymbol{\xi}^{(r)}}$ is independently sampled from the same distribution, a property which has also been used to obtain the final equality in Eq.~(B8). Inserting resolutions of the identity we obtain
\begin{align}
		\begin{split}
		&\ev{\delta\Theta_A^2}_M= \frac{1}{M}\sum_{n=1}^Z\sum_{m=1}^Z\sum_{n'=1}^Z\sum_{m'=1}^Z\mel{{n}}{\op{A}}{n'}\mel{m}{\op{A}}{m'} \\
		&\times\bigg(\ev{ {c_{n}^{(r)}}^*\! {c_{m}^{(r)}}^*\!c_{n'}^{(r)}\!c_{m'}^{(r)}}_M\!\!
		-\ev{{c_{n}^{(r)}}^* \!c_{n'}^{(r)}}_M\!\!\ev{{c_{m}^{(r)}}^* \!c_{m'}^{(r)}}_M\bigg).
	\end{split}
\end{align}
If $\op{A}$ is Hermitian, as it is in all of the cases we have considered in this paper, its eigenstates form a basis for the nuclear spin Hilbert space. By choosing the states $\ket{n}$ to be these eigenstates, such that $\op{A}\ket{n} = \ket{n}\alpha_n$, the above expression simplifies to
\begin{align}\label{delta-thetaA-1-eq}
	\begin{split}
		&\ev{\delta\Theta_A^2}_M = \frac{1}{M}\sum_{n=1}^Z\sum_{m=1}^Z\alpha_n\alpha_m \\
		&\times\bigg(\ev{|c_n^{(r)}|^2|c_m^{(r)}|^2}_M\!\!
		-\ev{|c_n^{(r)}|^2}_M\!\!\ev{|c_m^{(r)}|^2}_M\bigg).
	\end{split}
\end{align}
If the randomly sampled states $\ket*{\psi^{(r)}}$ are normalised to 1 (i.e. $\braket*{\psi^{(r)}}=1$), then $\sum_{n=1}^Z |c_n^{(r)}|^2 = 1$ and therefore
\begin{align}\label{corr-sum-eq}
	\sum_{n=1}^Z\bigg(\ev{|c_n^{(r)}|^2|c_m^{(r)}|^2}_M\!\!
	-\ev{|c_n^{(r)}|^2}_M\!\!\ev{|c_m^{(r)}|^2}_M\bigg) = 0,
\end{align}
and thus Eq.~\eqref{delta-thetaA-1-eq} can be re-written as
\begin{align}\label{delta-thetaA-2-eq}
	\begin{split}
		&\ev{\delta\Theta_A^2}_M = \frac{1}{M}\sum_{n=1}^Z\sum_{m=1}^Z\Delta\alpha_n\Delta\alpha_m \\
		&\times\bigg(\ev{|c_n^{(r)}|^2|c_m^{(r)}|^2}_M\!\!
		-\ev{|c_n^{(r)}|^2}_M\!\!\ev{|c_m^{(r)}|^2}_M\bigg).
	\end{split}
\end{align}
where $\Delta \alpha_n = \alpha_n - (1/Z)\sum_{m=1}^Z\alpha_m$. From this expression, we can obtain an upper bound on $\ev{\delta\Theta_A^2}_M$, from which we can derive conditions for the efficient convergence of a given sampling method.

First we note that $|\Delta\alpha_n|\leq \Delta_A$, where $\Delta_A$ is the range of eigenvalues of $\op{A}$. (For quantities like the singlet yield and the time-dependent singlet survival probability of a radical pair recombination reaction, $\Delta_A$ will be bounded above by 1.) This gives the following upper bound on $\ev{\delta\Theta_A^2}_M$,
\begin{align}
	\begin{split}
	&\ev{\delta\Theta_A^2}_M \leq \frac{\Delta_A^2}{M} \\
	\times\sum_{n=1}^Z&\sum_{m=1}^Z \bigg|\ev{|c_n^{(r)}|^2|c_m^{(r)}|^2}_M\!\!
	-\ev{|c_n^{(r)}|^2}_M\!\!\ev{|c_m^{(r)}|^2}_M\bigg|.
\end{split}
\end{align}
Rearranging Eq.~\eqref{corr-sum-eq} gives
\begin{align}
	\begin{split}
	&\ev{|c_m^{(r)}|^4}_M-\ev{|c_m^{(r)}|^2}_M^2 = \\
	&-\sum_{n\neq m}\bigg(\ev{|c_n^{(r)}|^2|c_m^{(r)}|^2}_M\!\!
	-\ev{|c_n^{(r)}|^2}_M\!\!\ev{|c_m^{(r)}|^2}_M\bigg)
	\end{split}
\end{align}
and therefore this upper bound can be written as
\begin{align}\label{delta-thetaA-ub-eq}
	\begin{split}
		&\ev{\delta\Theta_A^2}_M \leq \frac{2\Delta_A^2}{M} \sum_{m=1}^Z \bigg|\ev{|c_m^{(r)}|^4}_M\!\!
		-\ev{|c_m^{(r)}|^2}_M^2\bigg|.
	\end{split}
\end{align}
The trace sampling will be self-averaging if the sum on the right-hand side of this equation is $\mathcal{O}(1/Z)$.\cite{Weisse2006}

It is instructive to first consider what $\ev{\delta\Theta_A^2}_M$ will be if the randomly sampled states happen to be maximally coherent in the $\ket{n}$ basis. By this we simply mean that in the $\ket{n}$ basis, $c_n^{(r)} = e^{i\phi_n^{(r)}}/\sqrt{Z}$, where $\phi_n^{(r)}$ is a random phase factor. In this special case, $|c_{n}^{(r)}| = 1/\sqrt{Z}$, and therefore $\ev{\delta\Theta_A^2}_M = 0$.\cite{Iitaka2004} This is because in this case $\mu_A = \ev*{\op{A}}{\psi^{(r)}}$ for any $\ket*{\psi^{(r)}}$. This situation is of course very unlikely to occur, but it demonstrates that if the sampled state is, in a very loose sense, spread out evenly in Hilbert space in the eigenbasis of $\op{A}$, the fluctuations in $\Theta_A$ will become smaller and the convergence of the trace sampling will be more efficient. 

Next we note that \textit{if} the states $\ket*{\psi^{(r)}}$ are sufficiently spread out in Hilbert space, then we can expect
\begin{align}\label{self-av-cond-eq}
	\ev*{|c_m^{(r)}|^4}_M = \mathcal{O}(1/Z^2),
\end{align}
and if this is satisfied then the upper bound of $\ev{\delta\Theta_A^2}_M $ from Eq.~\eqref{delta-thetaA-ub-eq} will be $\mathcal{O}(\Delta_A^2/(MZ))$. Eq.~\eqref{self-av-cond-eq} is thus a sufficient condition for the state sampling to be self-averaging, i.e. for $\ev{\delta\Theta_A^2}_M^{1/2}  = \mathcal{O}(1/\sqrt{Z})$, and therefore exponentially convergent in the number of spins in the Hilbert space. This will occur in our case when the spins in the time evolved states $\ket{\Psi_{\sing,\boldsymbol{\xi}}(t)}$ become highly entangled. Showing rigorously when $\ev*{|c_m^{(r)}|^4}_M = \mathcal{O}(1/Z^2)$ is quite challenging, although we will give one example below of a sampling method for which Eq.~\eqref{self-av-cond-eq} holds true for \textit{any} basis states $\ket{n}$.

\subsection{Different sampling methods}

Two types of trace sampling have been used previously in the radical pair spin dynamics literature: spin coherent state $\ket{\boldsymbol{\Omega}}$ sampling,\cite{Lewis2016,Fay2017,Keens2020} and spin projection state $\ket{\vb{M}}$ sampling.\cite{Lewis2016,Keens2020} Other schemes have also been proposed in the condensed matter physics literature.\cite{Weisse2006,Silver1994} Here we consider sampling generalised $SU(Z)$ coherent states $\ket{\vb{Z}}$,\cite{Nemoto2000,Runeson2020} which are simply random normalised states in the full nuclear spin Hilbert space. 

In order to expose the limitations of methods like $\ket{\boldsymbol{\Omega}}$ sampling and $\ket{\vb{M}}$ sampling, let us suppose that the full nuclear spin Hilbert space $\pazocal{H}$ can be decomposed into a direct product of $\pazocal{H}_0$ and the rest of the space $\pazocal{H}_1$, with dimensions $Z_0$ and $Z_1$ respectively. We will consider the case where the operator $\op{A}$ can be decomposed into a term $\op{A}_0$ which only acts on $\pazocal{H}_0$, and a perturbation $\Delta \op{A}$ which acts on the full space,
\begin{align}
	\op{A} = \op{A}_0 + \Delta \op{A}.
\end{align}
Let us also suppose the random states $\ket*{\psi^{(r)}}$ can be decomposed into a direct product of a normalised state in $\pazocal{H}_0$, $\ket*{\psi_0^{(r)}}$, and a normalised state in $\pazocal{H}_1$, $\ket*{\psi_1^{(r)}}$, and that these are sampled independently. We can write the eigenstates of $\op{A}_0$ as $\ket{n}=\ket{i,j}=\ket{i_0}\otimes\ket{j_1}$ such that $\op{A}_0\ket{i,j}=\alpha_{0,i}\ket{i,j}$, where $\ket{i_0}$ is a basis state in $\pazocal{H}_0$ and $\ket{j_1}$ is a basis state in $\pazocal{H}_1$. Hence the coefficient $c^{(r)}_{n} = c^{(r)}_{i,j}$ can be decomposed into a product of independent coefficients $c^{(r)}_{i,j} = c^{(r)}_{0,i}c^{(r)}_{1,j}$.  When the perturbation $\Delta\op{A}$ is neglected, the sums in Eq.~\eqref{delta-thetaA-2-eq} reduce to sums over the $Z_0$ basis states in $\pazocal{H}_0$, and the sum in Eq.~\eqref{delta-thetaA-ub-eq} that gives an upper bound on $\ev*{\delta\Theta_A^2}_M$ can therefore only be $\mathcal{O}(1/Z_0)$ at best. 

This case arises when a subset of nuclear spins dominate the hyperfine coupling, and the remaining hyperfine coupled nuclei can be treated perturbatively. In this case, to zeroth order in the perturbation, which is a valid approximation at short times, an operator of the form in Eq.~\eqref{A-op-gen-eq} will only act on a subset of the nuclear spins (those whose states, along with the states of the two electron spins, are in $\pazocal{H}_0$). It follows that the variance of observables that depend on the electron spins will only converge at best as $\mathcal{O}(1/Z_0)$, for both projection state sampling and coherent state sampling. This suggests that these trace sampling methods may not always be optimal.

A similar argument can be applied when the decomposition of $\pazocal{H}$ is into a direct sum of subspaces, $\op{A}_0$ contains terms which act on different subspaces, and $\Delta\op{A}$ contains terms which connect these subspaces. In this case, the convergence of the $\op{A}_0$ term in $\ev{\delta\Theta_A^2}_M$ is also reduced if the random states $\ket*{\psi^{(r)}}$ are confined to single subspaces in the direct sum. It has previously been noted\cite{Lewis2016,Keens2020} that projection state ($\ket{\mathbf{M}}$) sampling is often less efficient than coherent state ($\ket{\boldsymbol{\Omega}}$) sampling. This is because the projection states are eigenstates of the total angular momentum projection operator $\op{J}_z$, and $\op{A}$ often (at least approximately) conserves $\op{J}_z$, so $\ket{\mathbf{M}}$ sampling is restricted to the lower dimensional eigenspaces of $\op{J}_z$. For example, in the case of the \ce{FAD^{$\bullet-$}-X^{$\bullet$}} radical pairs considered by Keens \& Kattnig,\cite{Keens2020} the dominant term in the spin Hamiltonian is the \ce{FAD^{$\bullet-$}} nitrogen hyperfine coupling, which has axial symmetry and therefore approximately commutes with $\op{J}_z$. This may explain the observation in Ref.~\onlinecite{Keens2020} that coherent state sampling is consistently (slightly) more efficient than projection state sampling for these radical pairs.

The situation for $SU(Z)$ coherent state $\ket{\vb{Z}}$ sampling is considerably more appealing. In this case, we can obtain a closed-form expression for the upper bound on $\ev{\delta\Theta_A^2}_M$ in Eq.~\eqref{delta-thetaA-ub-eq} that holds for any basis $\ket{n}$ and does not involve any assumptions about either the structure of $\pazocal{H}$ or the form of the (Hermitian) operator $\op{A}$. This follows because when $\ket{\vb{Z}}$ sampling is used, $\ev*{|c_m^{(r)}|^{2p}}_M$ is given by
\begin{align}
 \ev{|c_m^{(r)}|^{2p}}_M =  \int_{\mathbb{R}^Z} \dd{\vb{X}}\int_{\mathbb{R}^Z} \dd{\vb{Y}}\frac{\delta(|\vb{Z}|-1)}{\mathcal{S}_{2Z}}\left(X_m^2+Y_m^2\right)^p,
\end{align}
the integrals in which can be evaluated using the following general formula involving Gamma functions\cite{Stanislav2005}
\begin{align}
	\int_{\mathbb{R}^{2Z}}\dd{\vb{Z}} \delta(|\vb{Z}|-1)\prod_{k=1}^{2Z} Z_k^{2p_k} = 2 \frac{\prod_{k=1}^{2Z} \Gamma(p_k+1/2)}{\Gamma(Z+\sum_{k=1}^n p_k)}.
\end{align}
Using this formula to evaluate Eq.~(B17) with $p=0$, 1, and 2 gives $\mathcal{S}_{2Z}=2\Gamma(1/2)^{2Z}/\Gamma(Z)$, $\ev*{|c_m^{(r)}|^2}_M=1/Z$, and $\ev*{|c_m^{(r)}|^4}_M = 2/[Z(Z+1)]$, and substituting these results into Eq.~\eqref{delta-thetaA-ub-eq} gives
\begin{align}\label{delta-thetaA-ub-sun-eq}
	\begin{split}
		&\ev{\delta\Theta_A^2}_M \leq \frac{2\Delta_A^2}{M} \frac{Z-1}{Z(Z+1)} = \frac{2\Delta_A^2}{MZ}+\mathcal{O}\left(\frac{1}{Z^2}\right).
	\end{split}
\end{align}
Note that this hold for any basis $\ket{n}$, because the distribution from which the $\ket{\vb{Z}}$ states are sampled is invariant under a unitary transformation [$\delta(|\vb{Z}|-1) = \delta(|\vb{U Z}|-1)$, where $\vb{U}$ is an arbitrary unitary matrix].

From this we can conclude that sampling $\ket{\vb{Z}}$ states should always be self-averaging, even when the operator $\op{A}$ only acts on a lower dimensional subspace of $\pazocal{H}$, because the $\ket{\vb{Z}}$ states are on average close to maximally coherent in all bases. This analysis suggests that this type of sampling will generally outperform coherent state and projection state sampling.
 
\subsection{Numerical comparison}

\begin{figure}[t]
	\includegraphics[width=0.47\textwidth]{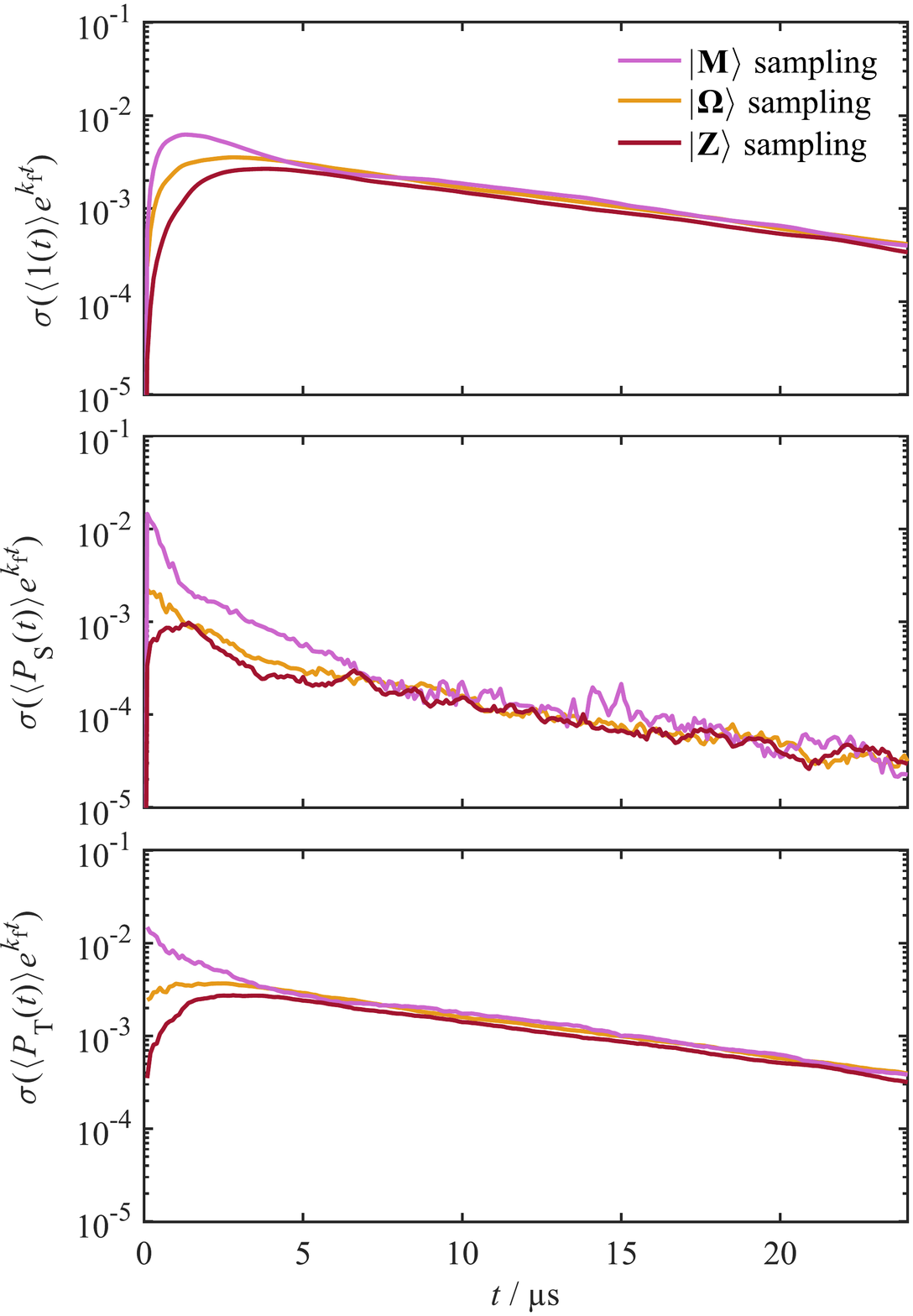}
	\caption{The standard errors in the means of various observables for the \ce{FAD^{$\bullet-$}-W^{$\bullet+$}} model with 12 hyperfine coupled spins, all computed with $M=128$  samples of the trace over nuclear spin states. The observables considered are $\ev{1(t)}e^{k_\mathrm{f}t}$ (top panel), $\ev{P_\sing(t)}e^{k_\mathrm{f}t}$ (middle panel) and $\ev{P_\trip(t)}e^{k_\mathrm{f}t}$ (bottom panel). In each case we see that $\ket{\vb{Z}}$ sampling outperforms the other sampling methods at short times, and that the entanglement caused by the spin dynamics improves the efficiency of the other sampling methods at longer times.} \label{se-methods-fig}
\end{figure}

Finally, to provide a numerical illustration of the relative efficiencies of the different sampling methods, we show in Fig.~4 the standard errors in the means of various observables for the \ce{FAD^{$\bullet-$}-W^{$\bullet+$}} model with 12 hyperfine coupled spins, as obtained using projection state $\ket{\vb{M}}$ sampling, spin coherent state $\ket{\boldsymbol{\Omega}}$ sampling, and $SU(Z)$ coherent state $\ket{\vb{Z}}$ sampling. As predicted by the above analysis, the standard errors are largest for $\ket{\vb{M}}$ state sampling, followed by $\ket{\boldsymbol{\Omega}}$ state sampling, with $\ket{\vb{Z}}$ state sampling significantly outperforming both at short times. However, after about 5 $\muup$s of evolution, once the dynamics has entangled the nuclear spins, the different sampling methods all perform similarly in terms of statistical convergence. This implies that even projection state sampling becomes self-averaging after a sufficiently long period of hyperfine-coupled spin evolution (for a system in which $\hat{J}_z$ is not rigorously conserved). 

In the calculations reported in the main part of this paper, we used spin coherent state sampling, which proved to be sufficient for our purposes. However, it is clear from the results in Fig.~4 that $\ket{\vb{Z}}$ sampling is distinctly better at short times, and from the analysis we have given above that it will become even better still for larger spin systems. Since it does not require any more effort than $\ket{\boldsymbol{\Omega}}$ sampling (in fact it is somewhat simpler to implement), we believe that $\ket{\vb{Z}}$ sampling should be the method of choice for trace evaluation in future applications of the SSE to spin dynamics.


\begin{thebibliography}{99}
\bibitem{footnote1}
Strictly speaking, the SLE is only exact in the limit where the spin interaction energies are much
smaller than $k_{\rm B}T$.
\bibitem{footnote2} 
There are situations in which one might also want to allow the Haberkorn reaction operator to depend on the stochastically fluctuating variables $\sX(t)$, for example to include effects such as diffusive re-encounters of free radicals in solution. This is not necessary in either of the example applications of the SSE considered in this paper, but it would only require a tiny modification of the present theory: simply replace $\hat{K}$ with $\op{K}(t) \equiv \op{K}(\sX(t))$ in Eq.~(12) and treat this in the same way as we have treated $\op{H}(t)$.
\bibitem{footnote3}
In the context of the SSE with trace sampling, the averaging over realisations of the fluctuating variables $\sX(t)$ is accomplished by (i) sampling a different random realisation of $\sX(0)$ for each random nuclear spin state $\ket{\psi(\boldsymbol{\xi})}$ that is used in the trace sampling, and (ii) evolving each $\sX(t)$ along a different stochastic trajectory.
\end{thebibliography}
\end{document}